\documentclass[12pt]{article}

\usepackage{scicite}

\usepackage{times}

\newenvironment{sciabstract}{%
\begin{quote} \bf}
{\end{quote}}

\newcounter{lastnote}

\title{Millisecond extragalactic radio bursts as magnetar flares. Comment on
``A population of fast radio bursts at cosmological distances'' by Thornton et al.}

\author
{Sergei B. Popov,$^{1\ast}$ Konstantin A. Postnov$^{1}$\\
\\
\normalsize{$^{1}$Sternberg Astronomical Institute, Lomonosov Moscow State University,}\\
\normalsize{Universitetski pr. 13, Moscow, 119991, Russia}\\
\normalsize{$^\ast$To whom correspondence should be addressed; E-mail:
sergepolar@gmail.com.}
}

\date{}

\begin{document} 


\baselineskip24pt


\maketitle 

\begin{sciabstract}
Properties of the population of millisecond extragalactic radio bursts
discovered by Thornton et al. (2013) are in good correspondence with
the hypothesis that such events are related to hyperflares of magnetars,
as was proposed by us after the first observation of an extragalactic millisecond
radio burst by Lorimer et al. (2007). We also point that some of multiple
millisecond radio bursts from
M31 discovered by Rubio-Herrera et al. (2013) also can be related to weaker
magnetar bursts.\footnote{This note is based on the Comment published on-line in
the Science magazine:
http://comments.sciencemag.org/content/10.1126/science.1236789\#comments. 
The note is submitted only to the arXiv, and
should be cited by its arXiv identifier.} 
\end{sciabstract}

Recently, a population of fast radio bursts (FRBs) of apparently
extragalactic origin has been discovered \cite{frb}.
Their nature  has been disputed since the first detection of a millisecond
extragalactic radio burst  \cite{lorimer}.
Apparently non-thermal, these bursts can be formed in relativistic plasma
which can be produced in a variety of astrophyical events (core-collapse supernovae, 
neutron star mergers, etc.). 
A natural explanation of FRBs  can also be
related to hyperflares from soft gamma-ray repeaters (SGRs), 
as was suggested in 
\cite{pp_merb}.

SGRs are thought to be strongly magnetized neutron stars,
which exhibit regular flaring activity, which is thought to be 
related to strong internal
magnetic field dissipation \cite{sandro,esposito}.
Energetic giant flares  can be due to  relativistic
magnetized explosions in SGR magnetospheres \cite{lyutikov2006},
and the appearence of
non-thermal radio emission associated with SGR giant flares and hyperflares 
seems natural. 
The possible physical mechanism for such radio bursts was
suggested in \cite{lyutikov2002}.  
Due to induced scattering, short radio bursts with high brightness
temperature can propagate only in relativistic plasmas  \cite{lyubarsk}.

Radio fluxes of FRBs agree with the  
scaling suggested by \cite{lyutikov2002}. A hyperflare
with $L=10^{47}$~erg~s$^{-1}$ (like the one detected from SGR 1806-20 on
Dec. 27, 2004, see \cite{bork, hurley, palmer}) 
at a distance of 600 Mpc could produce a FRB
with a flux $\sim30$~Jy \cite{pp_merb}. FRBs detected by \cite{frb} are
more distant, so their radio fluxes should to be correspondingly lower, as
observed. 

 The rate of hyperflares for extragalactic SGRs, and so for millisecond
extragalactic radio bursts, was estimated in
\cite{pp_merb} to be 20-100~d$^{-1}$~Gpc$^{-3}$. This   
is confirmed by new FRB observations \cite{frb} 
giving the rate $\sim 10^4$~d$^{-1}$ from the whole sky up to
a distance of $\sim 3$~Gpc. Thus,
new observations \cite{frb} 
support the model of FRB origin due to SGR hyperflares from the point of
view of predicted fluxes and statistics.

Large dispersion measure, $\sim500$-1000, detected for FRBs \cite{frb} is
not unexpected in the case of SGR flares, as these sources are associated
with 
recent massive star formation \cite{popov}. Galaxies at $z\sim0.5$-1,
in which FRBs were discovered, have significant massive star formation 
and are expected to be
rich in gas and dust,  resulting in large dispersion measure.

We also note that 
short millisecond radio burts with fluxes $\sim1$--4 Jy from M31 have recently been
reported \cite{rubio}. Some of these events, especially multiple bursts with
similar dispersion measures, can be connected with the repetitive 
activity of magnetars in M31. Candidates to hyperflares of magnetars in M31 have
already been reported \cite{mazets}. 
In the present case not hyperflares, but weaker SGR bursts with
luminosities $L\sim 10^{39}$~--~$10^{40}$~erg~s$^{-1}$ can be responsible for
the observed radio bursts, assuming the SGR burst energy scaling from
\cite{lyutikov2002}.  

The rate of bursts obtained in \cite{rubio} 
is in agreement with that of
X/gamma-ray bursts from magnetars 
with required luminosity assuming one SGR in M31 
to be in active phase --- a few events per hour. Such a rate
was observed by \cite{gogus} during the RXTE monitoring of a Galactic SGR
during its active phase. 
X-ray (or/and gamma-ray) bursts with $L\sim
10^{39}$~--~$10^{40}$~erg~s$^{-1}$ by themselves
can be easily missed from M31, because of their short duration. Despite
potentially detectable fluxes, the number of photons from $\sim0.1$~s burst
would be too low to be detected and identified as a real source.

However,
if multiple radio bursts in M31 are related to magnetars activity, their
rate is expected to fluctuate significantly.  Future observations 
can be used to check the expected radio burst energy distribution 
$dN/dE\sim E^{-1.66}$ known for SGR bursts \cite{gogus}. 


\bibliography{pp}

\bibliographystyle{Science}

\end{document}